%Version as published in Phys. Lett. B, but with minor errors corrected, 
%in particular f(T)-->f(0).
%Please select the Harvmac ``b'' option.
%\overfullrule 0pt 
\input harvmac
\input epsf

%Askitest
% Upper-case    A B C D E F G H I J K L M N O P Q R S T U V W X Y Z
% Lower-case    a b c d e f g h i j k l m n o p q r s t u v w x y z
% Digits        0 1 2 3 4 5 6 7 8 9
% Exclamation   !           Double quote "          Hash (number) #
% Dollar        $           Percent      %          Ampersand     &
% Acute accent  '           Left paren   (          Right paren   )
% Asterisk      *           Plus         +          Comma         ,
% Minus         -           Point        .          Solidus       /
% Colon         :           Semicolon    ;          Less than     <
% Equals        =           Greater than >          Question mark ?
% At            @           Left bracket [          Backslash     \
% Right bracket ]           Circumflex   ^          Underscore    _
% Grave accent  `           Left brace   {          Vertical bar  |
% Right brace   }           Tilde        ~
%%%%%%%%%%%%%%%%%%%%%%%%%%%%%%%%%%%

\def\half{{\textstyle{1\over2}}}
\def\ga{\gamma}

\def \la{\lambda}

\def\frak#1#2{{\textstyle{{#1}\over{#2}}}}

\def\tbar{{\overline{t}}}

\def\ybar{{\overline{y}}}

\def\sic{supersymmetric}

\def\ssm{supersymmetric standard model}

\def \in{\leftskip = 40 pt\rightskip = 40pt}
\def \out{\leftskip = 0 pt\rightskip = 0pt}

\def\npb{{Nucl.\ Phys.\ }{\bf B}}
\def\physrep{Phys.\ Reports\ }
\def\plb{{Phys.\ Lett.\ }{\bf B}}

\def\prb{{Phys.\ Rev.\ }{\bf B}}
\def\prd{{Phys.\ Rev.\ }{\bf D}}

\def\lf{16\pi^2}
\def\llf{(16\pi^2)^2}
\def\lllf{(16\pi^2)^3}
\def\llllf{(16\pi^2)^4}

{\nopagenumbers
\line{\hfil LTH 386}
\line{\hfil hep-ph/9610296}
\line{\hfil Revised Version}
\vskip .5in
\centerline{\titlefont The Quasi-Infra-Red Fixed Point at Higher Loops}
\vskip 1in
\centerline{\bf P.M.~Ferreira, I.~Jack and D.R.T.~Jones}
\bigskip
\centerline{\it DAMTP, University of Liverpool, Liverpool L69 3BX, U.K.}
\vskip .3in
We calculate the four-loop $\beta$-function for the generalised 
Wess-Zumino model. We use the result and Pad\'e-Borel 
summation to discuss the domain of 
attraction of the quasi-infra-red fixed point of the top-quark 
Yukawa coupling in the supersymmetric standard model, and argue 
that the domain  is in fact substantial.

\Date{October 1996}}

In this note we present the four-loop $\beta$-function 
for a many-field Wess-Zumino model with arbitrary cubic interactions. 
(This calculation is a straightforward generalisation of the existing
one of  Avdeev et al\ref\avdeev{L.V.~Avdeev et al,  \plb 117 (1982) 321}
for the single field and coupling case).  We use the result to write
down the $O(\lambda_t^9)$ terms in the $\beta$-function  for the
top-quark Yukawa coupling in the \ssm. We then use this result  to
discuss issues regarding the limitations of perturbation theory in  the
standard running analysis whereby the low energy \ssm\ is  matched onto
a much reduced (in terms of the number of free  parameters) theory at
high energies. In such analyses, putative  \sic\ spectra are often
presented as scatter-plots, for ranges of certain of the free
parameters, subject to certain cuts.  One such cut, frequently applied,
is that the Yukawa couplings remain  perturbative throughout: in for
example Ref.~\ref\pierce{D.M.~Pierce et al, hep-ph/9606211}\  one finds
that  $\lambda_{t,b,\tau} \leq 3.5$  has been enforced. What is the
basis and  reliability of such cuts, given the  (presumably) asymptotic
nature of the perturbation expansion?  This question becomes
particularly interesting in the context of  the possible 
quasi-infra-red fixed point (QIRFP) behaviour of $\la_t$
\ref\Pendleton{B.~Pendleton and G.G.~Ross,
\plb 98 (1981) 291}%
\nref\hill{C.T.~Hill, \prd 24 (1981) 691}%
--\ref\lanz{C.T.~Hill,
C.N.~Leung and S.~Rao, \npb 262 (1985) 517\semi
C.D.~Froggatt, R.G.~Moorhouse and I.G.~Knowles, \plb 298 (1993) 356\semi 
M.~Carena et al, \npb 419 (1994) 213\semi
W.A.~Bardeen et al, \plb 320 (1994) 110\semi
M.~Lanzagorta and G.G.~Ross, \plb 349 (1995) 319}.  
 
We begin with the  generalised Wess-Zumino model, 
defined by the superpotential:
\eqn\Ea{
W=\frak{1}{6}Y^{ijk}\Phi_i\Phi_j\Phi_k.}
where $\Phi_i$ is a multiplet of chiral superfields. 
The $\beta$-functions for the Yukawa couplings $\beta_Y^{ijk}$
are given by
\eqn\Ec{
\beta_Y^{ijk}= Y^{p(ij}\ga^{k)}{}_p =
Y^{ijp}\ga^k{}_p+(k\leftrightarrow i)+(k\leftrightarrow j),}
where $\ga$ is the anomalous dimension for $\Phi$.

The result for $\ga$ through three loops is as follows
\ref\jjnb{I.~Jack, D.R.T.~Jones and C.G.~North, \npb 473 (1996) 308}: 
\eqna\Ed$$\eqalignno{
\lf\ga^{(1)}&=P &\Ed a\cr 
\llf\ga^{(2)}&=-S_4&\Ed b\cr
\lllf\ga^{(3)} &= 
\frak{3}{2}\zeta(3)M  +2Y^*S_4 Y - \half S_7 - S_8&\Ed c\cr}$$
where our notation follows that of 
Ref.~\ref\jjn{I.~Jack, D.R.T.~Jones and C.G.~North, hep-ph/9609325}, 
except that here we have no gauge coupling:
\eqna\tlfb$$\eqalignno{
P^i{}_j&=\half Y^{ikl}Y_{jkl}&\tlfb a\cr
S_4 ^i{}_j &= Y^{imn}P^p{}_m Y_{jpn}&\tlfb b\cr
(Y^*S_4 Y)^i{}_j &= Y^{imn}S_4{}^p{}_m Y_{jpn}.&\tlfb c\cr
S_7 ^i{}_j &= Y^{imn}P^p{}_m P^q{}_n Y_{jpq}&\tlfb d\cr
S_8 ^i{}_j &= Y^{imn}(P^2)^p{}_m Y_{jpn}&\tlfb e\cr
M^i{}_j &= Y^{ikl}Y_{kmn}Y_{lrs}Y^{pmr}Y^{qns}Y_{jpq}.&\tlfb f\cr
}$$
Our result for $\ga^{(4)}$ is as follows:
\eqn\gafour{\eqalign{
\llllf\ga^{(4)} &= \frak{5}{6}Y^{ikm}S_7{}^l{}_kY_{jlm} 
+(\frak{5}{3}-2\zeta(3))Y^{ikm}S_8{}^l{}_kY_{jlm}
+\frak{4}{3}Y^{ikm}(PS_4 + S_4 P)^l{}_kY_{jlm}
\cr& 
- 5Y^{ikq}Y_{kmp}S_4{}^n{}_mY^{lnp}Y_{jlm}
-(\frak{3}{2}\zeta(3) + \frak{3}{4}\zeta(4))Y^{ikm}M{}^l{}_kY_{jlm} 
\cr&
+(2\zeta(3)-1)\left[Y^{ikm}(P^3)^l{}_kY_{jlm} 
+Y^{ikm}(P^2)^l{}_k P^n{}_m Y_{jln}\right]\cr 
&+\frak{4}{3}Y^{ikm}P^l{}_k (S_4)^n{}_m Y_{jln}
-10\zeta(5)Y^{ikl}Y_{kmn}Y_{lsv}Y^{nsu}Y_{rtu}Y^{pmr}Y^{qtv}Y_{jpq}
\cr&-(6\zeta(3) -3\zeta(4))
[\half Y^{ikt}P^l{}_t Y_{kmn}Y_{lrs}Y^{pmr}Y^{qns}Y_{jpq}\cr&
+ \half Y^{ikl}Y_{kmn}Y_{lrs}Y^{pmr}Y^{nst}P^q{}_t Y_{jpq}
+ Y^{ikl}Y_{kmt}P^t{}_nY_{lrs}Y^{pmr}Y^{qns}Y_{jpq}
].\cr
}}        
We have explicitly calculated the requisite Feynman diagrams for the most part.
(This was in any case necessary since in Ref.~\avdeev\ results are 
given for subsets of, not individual graphs.) The results for several of
the Feynman integrals are given in Ref.~\jjn. In two places, 
however, we have relied on previous authority: specifically, in 
Table 1 of Ref.~\avdeev, we have used their total for the set of 
graphs $4.1$ to finesse the calculation of one particular graph; 
and we have used the result of Ref.~\ref\kazz{K.G.Chetyrkin et al, 
\npb 174 (1980) 345\semi
D.I.~Kazakov, \plb 133 (1983) 406}\ 
for the Feynman integral that arises 
in the calculation of the graph $4.5$.  

For the  \ssm\ superpotential, which is (retaining only $\la_t$) 
\eqn\smw{W = \la_tH_2 Q\tbar}
where $Q = \pmatrix{t\cr b\cr},$ 
we find, using $\beta_{\la_t} = (\ga_{H_2} + \ga_Q + \ga_{\tbar})\la_t$ 
\ref\thlp{P.M. Ferreira, I. Jack and D.R.T. Jones, \plb387 (1996) 80}:

\eqna\Kevd$$\eqalignno{
\lf\beta_{\lambda_t}^{(1)} &= 6\lambda_t^3
&\Kevd a\cr
\llf\beta_{\lambda_t}^{(2)} &= -22\lambda_t^5
&\Kevd b\cr
\lllf\beta_{\lambda_t}^{(3)} &= [102+ 36\zeta(3)]\lambda_t^7
&\Kevd c\cr}$$
and finally from Eq.~\gafour:
\eqn\bfour{\llllf\beta_{\lambda_t}^{(4)}=
-\left[678+696\zeta(3) -216\zeta(4)+1440\zeta(5)\right]\lambda_t^9.}

A specific aspect of the running analysis that  has become popular    
in recent years is the possibility that $\lambda_t$
may  exhibit QIRFP behaviour. The main
attraction  of this philosophy is the idea that a large range of  input
couplings at high energies may produce the same value of  $\lambda_t$ at
$M_Z$. It is clearly interesting to inquire as to the extent to which
this range is limited by the requirement of perturbative believability. 
Let us review the QIRFP paradigm:   the one-loop equations governing the
evolution of the $\lambda_t$ and the gauge couplings are 
\eqn\ssmad{
{{dy_t}\over{dt}} = y_t (6y_t  - \frak{16}{3}\alpha_3 -
3\alpha_2 - \frak{13}{15}\alpha_1), 
\quad\hbox{and}\quad 
{{d\alpha_i}\over{dt}}= b_i\alpha_i^2  
}
where  
$y_t=\frak{1}{4\pi}\lambda_t^2$, $\alpha_i = \alpha_{1,2,3}$,
$b_i = \pmatrix{\frak{33}{5} & 1 & -3\cr}$, 
 and $t = {1\over{2\pi}}\ln\mu$. 
We have made the approximation 
that $y_t >> y_\tau, y_b$,which involves the assumption 
that it is not the case that $\tan\beta >> 1$; 
it is straightforward to consider 
the more general case but we omit it here for simplicity. 
From Eq.~\ssmad\ we obtain:
\eqn\ssmak{
y_t (M_X ) = {{y_t (M_Z)}\over{f(0) - 6Fy_t (M_Z)}}  
}
where
\eqn\ssmai{F = \int_0^{T} f(\tau)\,d\tau, \quad 
T = {{1}\over{2\pi}}\ln{{M_X}\over{M_Z}}}

and 
\eqn\ssmaj{
f(\tau) = \left[{{\alpha_3 (T) }
\over{\alpha_3 (\tau)}}\right]^{16\over{3b_3}}
\left[{{\alpha_2 (T) }\over{\alpha_2 (\tau)}}\right]^{3\over{b_2}} 
\left[{{\alpha_1 (T) }\over{\alpha_1 (\tau)}}\right]^{13\over{15b_1}}.
}
We will take $M_X$ to be the unification scale,
requiring that $\alpha_2 (T) = \alpha_1 (T)$.  
Of course it is a straightforward matter to integrate the differential
equations  numerically, but the partial analytic solution above is
nevertheless useful  in order to see what is going on. 
From Eq.~\ssmak\ we see that 
$y_t$ suffers  a Landau pole unless 
\eqn\ssmal{
y_t (M_Z) < f(0)/(6F).
}
Another nice way of thinking about this result is as follows. 
Suppose 
\eqn\ssmam{
y_t(M_X) >> 1/(6F).
}
Then it is easy to see from Eq.~\ssmak\ that in this limit  
\eqn\ssman{
y_t (M_Z) = \ybar_t = f(0)/(6F)
}
which is independent of $y_t(M_X)$! Of course $\ybar_t$ is the 
value of $y_t (M_Z)$ such that the Landau pole occurs at $M_X$, 
and hence represents an upper limit; but it is more productive 
to think of $y_t (M_X)$ as the input, as follows.  
Depending on the value of $F$, there may be  a wide 
range of values of $y_t (M_X)$ which all lead to the same value of 
$y_t(M_Z)$. The corresponding value, $y_t (M_Z) = f(0)/(6F)$,     
is called a {\it quasi-infra-red fixed point} of the evolution 
equations. The prefix {\it quasi-}\ is occasioned by the fact that 
$f/(6F)$ is of course a function of $T$. 
This behaviour is illustrated in Fig.~1. 

\epsfysize= 2.5in
\centerline{\epsfbox{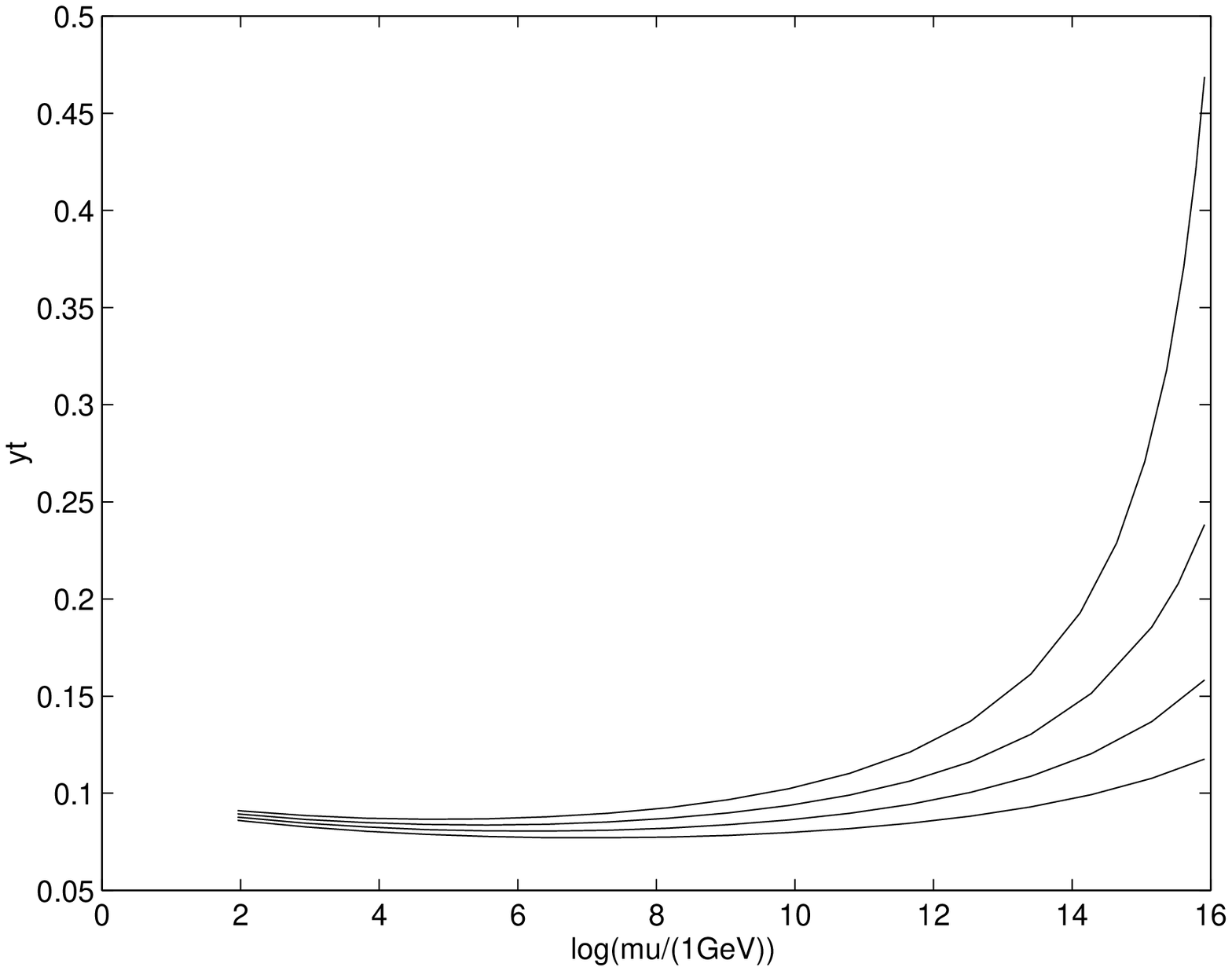}}
\in
{\it \noindent Fig.1:
Plot of $y_t$ against $\log_{10}{\mu\over{1{\rm GeV}}}$ for various values of
$y_t(M_X)$}
\medskip
\out
If we take the limit 
$T\to\infty$, then we approach  genuine fixed point behaviour
\Pendleton. It is easy to see this in the approximation 
$\alpha_1 = \alpha_2 = 0$, for which the equations for 
${{dy_t}\over{dt}}$ and ${{d\alpha_3}\over{dt}}$ exhibit 
an infra-red fixed point such that
$$
{y_t \over \alpha_3} = {{-3 + \frak{16}{3}}\over{6}} = 0.39.
$$
For $\alpha_3 (M_Z) = 0.1$ this gives 
$y_t (M_Z) = 0.039$. Substituting in Eq.~\ssmai\ and \ssmaj\ (with 
$\alpha_1(M_Z)=0.0167$, $\alpha_2(M_Z)=0.0320$)
one finds 
$f(0) = 9.98$ and $F=17.88$ so that $\ybar_t = 0.093$. Thus 
QIRFP and IRFP behaviour are quite distinct; it is the 
former which is relevant for the \ssm. (This was
first pointed out for the standard model in Ref.~\hill.) 

We now see that since $1/(6F)\approx 0.01$ we have that there is a wide range 
for $y_t(M_X)$ such that 
\eqn\lims{1/(6F) << y_t(M_X) < 1}
where the upper limit is a naive constraint  for  perturbative
believability.  (Note that the constraint $\lambda < 3.5$ used in 
Ref.~\pierce\  would correspond to $y_t < 0.975$).  We want to address
this latter restriction; does our four-loop calculation  afford any
insight into it?   In the literature, cuts of the type $\la_t \leq 3.5$
alluded to above are  sometimes motivated by requiring $\beta^{(2)} \leq
x\beta^{(1)}$, where  $\beta^{(L)}$ is the appropriate $L$-loop
$\beta$-function, and  $x$ is some convincing fraction: $\frak{1}{4}$
for example.  This leads,  using Eq.~\Kevd{}\  to $y_t \leq 0.86$. A
naive  extension of this approach would obviously  suggest a drastic
curtailment of the  acceptable range of $y_t (M_X)$, since for instance 
$\beta^{(4)} \leq \frak{1}{4}\beta^{(3)}$ gives $y_t \leq 0.16$ \foot{
Such naive constraints may be even more restrictive in 
other sectors of the theory: in particular the soft supersymmetry breaking
sector. By demanding that $\beta^{(2)}_{m^2_Q} \leq 
\frak{1}{4}\beta^{(1)}_{m^2_Q}$, we get 
\ref\tlss{I. Jack and D.R.T. Jones, \plb 333 (1994) 372\semi
S.P.~Martin and M.T.~Vaughn, \prd50 (1994) 2282\semi
Y.~Yamada, \prd50 (1994) 3537\semi
I.~Jack, et al,  \prd 50 (1994) R5481} $y_t \leq 0.32$, 
for any theory with the commonly 
assumed universal form of the soft supersymmetry breaking parameters 
(specific models
might originate smaller upper bounds; the ``$P = \frak{1}{3} Q$'' class of 
models
\ref\pthq{I.~Jack and D.R.T.~Jones, \plb 349 (1995) 294}, for example, gives
$y_t \leq 0.21$). 
}. This is
 illustrated in Fig.~2, where we plot $y_t (M_X)$ against $y_t (M_Z)$ 
for $L =1,\cdots 4$. 

\epsfysize= 2.5in
\centerline{\epsfbox{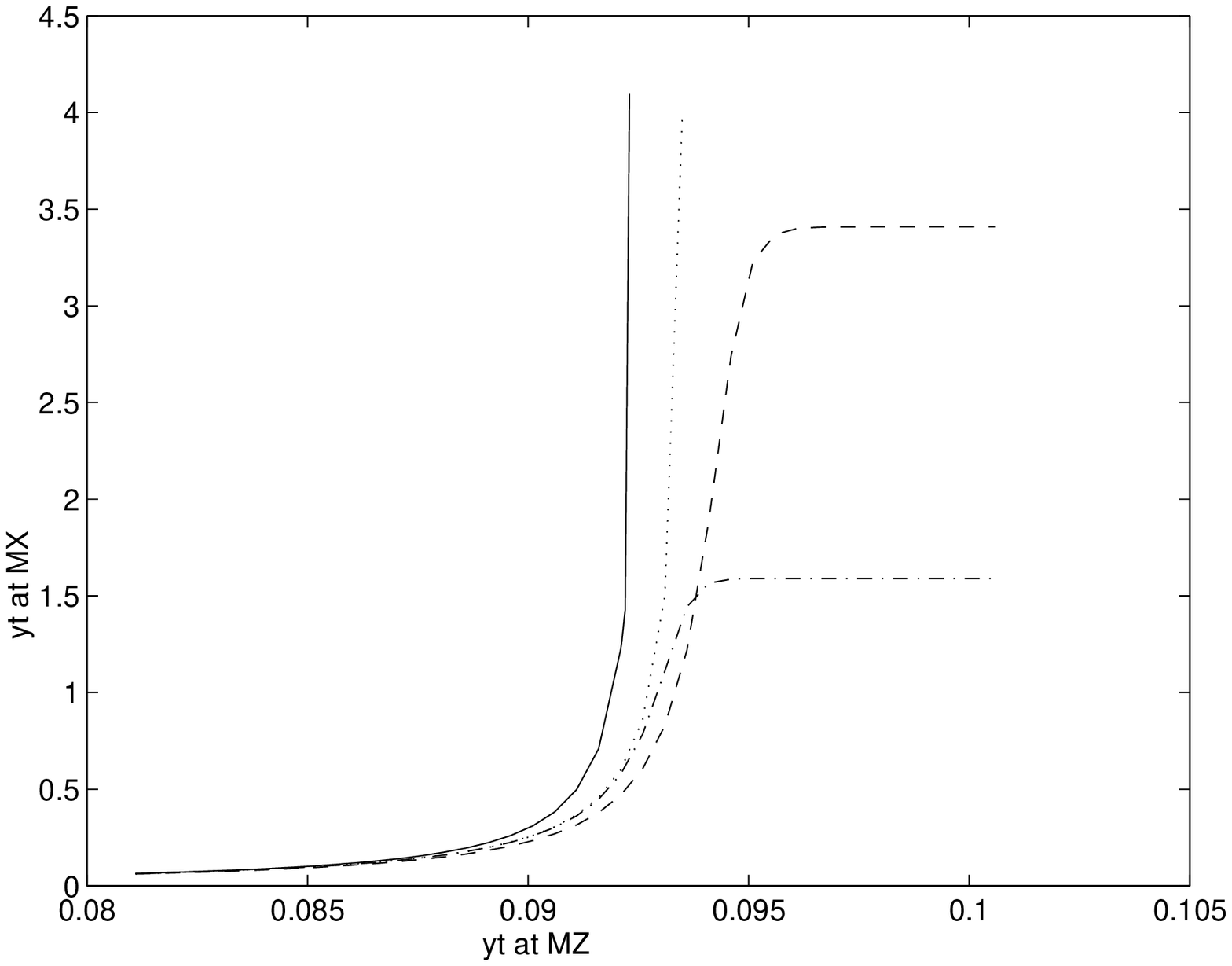}}
\in
{\it \noindent Fig.2:
Plot of $y_t(M_X)$ against $y_t(M_Z)$.
The solid, dashed, dotted and dash-dotted lines correspond to one, two, three
and four-loop $\beta$-functions respectively.}
\medskip
\out
%The equivalent four-loop restriction would very likely be
%more restrictive than the one coming from the Yukawa beta function. 
%One might
%ask if a perturbative limit coming from the soft parameter sector should have
%any relevance to the Yukawa sector. The answer is, not directly, but the 
%running of the soft parameters is vital to calculate threshold corrections,
%which determine the scale above which the Standard Model RGE's should be
%replaced by the SSM ones.

We wish to argue that the (presumably) asymptotic
nature of the  perturbation series for the $\beta$-functions means that
in fact the actual domain  of attraction of the QIRFP is more accurately
represented by the one-loop than by  the four-loop approximation.

A striking feature of Eq.~\bfour\ is  the broad similarity to the
corresponding result \ref\kaz{D.I. Kazakov, O.V. Tarasov and  A.A.
Vladimirov,   Sov.Phys. JETP 50 (1979) 521} for $O(n)$ $\phi^4$. Most 
importantly, note the characteristic alternating sign  behaviour,
suggesting the possibility of Borel summability.  (For a review on the
problem of resummation of perturbation theory see 
Ref.~\ref\zinn{J.~Zinn--Justin, \physrep 70 (1981) 110}.) We have not
found in the  literature any discussion of the large order behaviour of
the Wess-Zumino  model; but for the \sic\ anharmonic oscillator, it has
been noted \ref\verb{J.J.M. Verbaarschot and P. West, \prd 42 (1990)
1276; {\it ibid\/}  {\bf D}43 (1991) 2718}  that  while the \sic\ case
does represent a bifurcation point with respect to  some behaviour,
nevertheless the characteristic factorial divergence $L!$  persists.
It is interesting to note, however, that the exact form 
for the gauge $\beta$-function in an $N=1$, $d=4$ theory without 
chiral fields,
\ref\tim{D.R.T.~Jones, \plb 123 (1983) 45\semi
M.T.~Grisaru and P.~West, \npb 254 (1985) 249}
\ref\nov{V.~Novikov et al, \npb 229 (1983) 381\semi
V.~Novikov et al, \plb166 (1986) 329\semi
M.~Shifman and A.~Vainstein, \npb 277 (1986) 456}
\eqn\russa{\beta_g =
{{g^3}\over{\lf}}\left[ {{-3C(G)}
\over{1- 2C(G)g^2{(\lf)}^{-1}}}\right],}
clearly has a finite radius of convergence. Now this result may not 
hold in DRED (dimensional 
reduction with minimal subtraction); it was shown in 
Ref.~\ref\jjn{I.~Jack, D.R.T.~Jones and C.G.~North, \plb386 (1996) 138}
that the scheme in which Eq.~\russa\ is valid differs from  DRED.  
In any case, for the Wess-Zumino model, there is certainly no indication from 
Eq.~\gafour\ that the Yukawa $\beta$-function is other than an 
asymptotic series.\foot{Note that if this was true the same would hold for 
the exact result of Ref.~\nov\ for $\beta_g$, when chiral fields are present.}  
Now the existence of renormalon singularities  \ref\lautrup{B.~Lautrup,
\plb 69 (1977) 109}\ implies that in   asymptotically un-free theories
(such as the Wess-Zumino model)  amplitudes are not Borel-summable in
general; but  specific renormalisation group functions  may be so. It
seemed to us  worthwhile exploring the consequences of  naive
Pad\'e-Borel (PB) summation of the $y_t$-dependence of $\beta_{y_t}$. 

PB summation proceeds as follows (see for example 
Ref.~\ref\bernie{G.A.~Baker Jr., B.G.~Nickel and D.I.~Meiron, 
\prb 17 (1978) 1365\semi
J.~Ellis et al, \plb 366 (1996) 268}). Given a series
\eqn\pbone{f(x) = \sum_{n=0}a_n x^n}
one defines 
\eqn\pbtwo{B(x) = \sum_{n=0}a_n {{x^n}\over{n!}}}
and calculates $[N, M]$ Pad\'e approximants to $B(x)$, $B_{NM}$.   Then
the PB-summed version of $f(x)$ is given by 
\eqn\pbthree{F_{NM}(x) =
\int_0^{\infty} e^{-t}B_{NM} (xt)\, dt.} 
Essentially this construction amounts to a guess of the coefficients  of
powers of $x$ beyond those originally calculated, incorporating  (for
$x^L$) a  factor $L!$.   

We have calculated the $[1,1]$ and the $[2,2]$ Pad\'es for the $t$-Yukawa 
$\beta$-function, i.e. the series 
\eqn\pade{\eqalign{{{dy_t}\over{dt}} &= y_t\Biggl\{
 0 + 6y_t -\frak{1}{{4\pi}}22y_t^2
+(\frak{1}{{4\pi}})^2\left[102 + 36\zeta(3)\right]y_t^3\cr
&-(\frak{1}{{4\pi}})^3\left[678+696\zeta(3) -216\zeta(4)+1440\zeta(5)\right]
y^4_t+\cdots\Biggr\}\cr}} 
We have written the series in this form in order that the $O(y_t^L)$ term in 
the curly brackets should correspond to the $L$th order contribution in
perturbation theory. The $L!$ factor produced by the PB process for the 
$O(y_t^L)$ term in the series then correctly mimics the expected $L!$ growth 
of the $L$-loop perturbation theory contributions.
The $[1,1]$ PB approximation for the series
\eqn\padea{
{{dy_t}\over{dt}}=y_t(0+ay_t-by_t^2)}
is easily calculated as
\eqn\padeb{
{{dy_t}\over{dt}}=y_t\left[{{2a^2}\over{b}}
-{{4a^3}\over{b^2y_t}}e^{{2a}\over{by_t}}
{\rm E_1}\left({2a\over{by_t}}
\right)\right],}
where ${\rm E_1}(x) = \int_x^{\infty} {e^{-t}\over{t}}dt$ 
is the exponential integral function. 
%The $[1,2]$ approximation for the series
%\eqn\padebc{
%{{dy_t}\over{dt}}=y_t(0+ay_t+by_t^2+cy_t^3)}
%is given by 
%\eqn\padebd{
%{{dy_t}\over{dt}}=y_t\left[{{(2ac-3b^2)y_t}\over 2c}-{9b^3\over 2c^2}\left\{
%1+{3b\over{2cy_t}}e^{-{3b\over{2cy_t}}}{\rm E_1}
%\left(-{3b\over{2cy_t}}\right)
%\right\}\right].}
The $[2,2]$ PB approximation for the series
\eqn\padebb{
{{dy_t}\over{dt}}=y_t(0+ay_t-by_t^2+cy_t^3-dy_t^4)}
is found to be
\eqn\padec{
{{dy_t}\over{dt}}=y_t\Biggr[\sigma r_1r_2+{{r_1^2r_2(\sigma r_1
- a )}\over
{(r_1-r_2)y_t}}e^{-{r_1\over{y_t}}}{\rm E_1}\left(-{r_1\over{y_t}}\right)
+{{r_1r_2^2(\sigma r_2- a )}\over 
{(r_2-r_1)y_t}}e^{-{r_2\over{y_t}}}{\rm E_1}\left(-{r_2\over{y_t}}\right)
\Biggr],}
where
\eqn\paded{
\sigma={{a^2d+3b^3 - 4abc}\over
{4ac-6b^2}},}
and $r_1$ and $r_2$ are the roots of
\eqn\padee{
(3bd-4c^2)r^2+(6ad-12bc)r+12(2ac-3b^2)=0.}
In each case the PB approximation involves the exponential integral
function; this is a generic feature. 
Note that with $a\ldots d$ as given in Eq.~\pade, 
the roots of Eq.~\padee\ are both  negative 
and real. It follows that in Eq.~\padec, no singularity is encountered in 
the exponential integrals; the same is true of   Eq.~\padeb. 
In the $[1,2]$ PB, however,  the roots
of the analogous quadratic have opposite signs. Now such a pole 
may represent a physical renormalon singularity; but since 
we do not encounter it in the $[2,2]$ Pad\'e, and because we have no 
specific knowledge of the asymptotic behaviour for this theory, we choose to 
ignore this  case. 

We have investigated the running of $y_t$ and $\alpha_{1,2,3}$ by adding 
the perturbative contributions involving the gauge couplings to
Eq.~\Kevd{}\ and evolving the gauge couplings using the perturbative
$\beta$-functions. This seems justified on the grounds that
$\alpha_{1,2,3}$ remain small over the range of interest. The results are
displayed in Fig.~3, where we plot $y_t(M_X)$ against $y_t(M_Z)$ for our
$[1,1]$ PB and $[2,2]$ PB approximations. For purposes of comparison, we
also give the  results obtained using the one-loop and four-loop
$\beta$-functions.  (Since we only know explicitly the pure Yukawa
contribution at the four-loop level, we must omit the gauge
contributions at this level; but in the region of  interest we know that
the Yukawa contributions are dominant.)  

\epsfysize= 2.5in
\centerline{\epsfbox{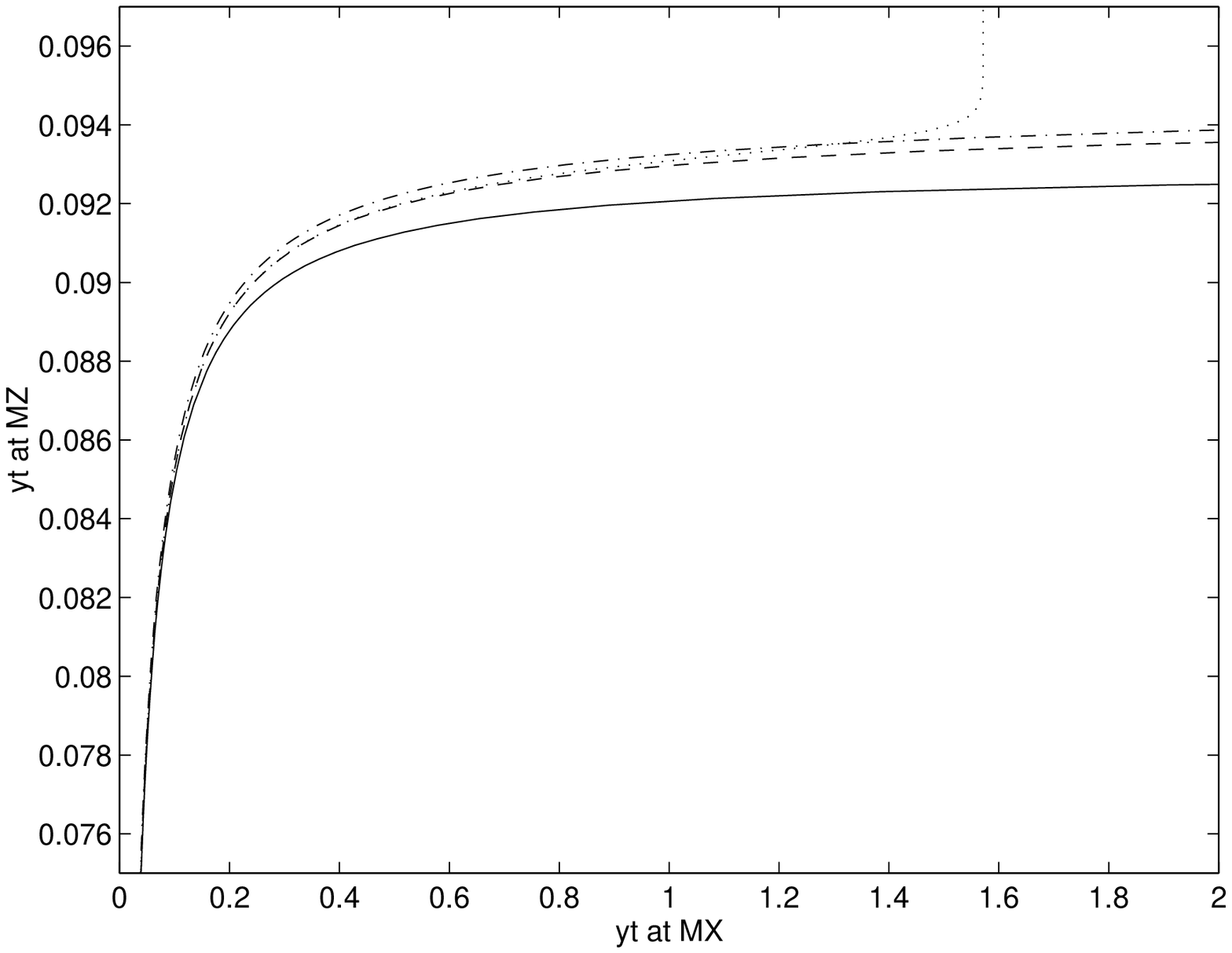}}
\in
{\it \noindent Fig.3:
Plot of $y_t(M_Z)$ against $y_t(M_X)$.
The solid and dotted lines correspond to the one and
four-loop perturbative $\beta$-functions, and the dot-dashed and dashed
lines to the $[1,1]$ and $[2,2]$ PBs respectively.}
\medskip
\out
For small $y_t$, the $[1,1]$ and $[2,2]$ PB results should approach the 
two-loop and four-loop perturbative results respectively; this is clearly seen
in the $[2,2]$ case. For large $y_t$, the asymptotic nature of the
perturbation series implies that lower orders in perturbation theory should
be more accurate than higher orders; accordingly, we see that as $y_t$ 
increases, the PB evolution starts to resemble the one-loop perturbative 
behaviour. Furthermore, the $[2,2]$ PB is closer to the one-loop result than
the $[1,1]$. We have also repeated this exercise for the $[2,1]$ PB; the
results are almost indistinguishable from those for the $[2,2]$ Pad\'e and
indicate that the successive PBs may be converging quite rapidly in the region
of interest. 

The one-loop perturbative evolution displays the QIRFP behaviour as explained 
in our earlier analysis; 
we see, for instance that values of $y_t(M_X)$ in the range
$0.2<y_t(M_X)<1$ lead to values of $y_t(M_Z)$ in the range 
$0.089<y_t(M_Z)<0.092$.
The upper limit on $y_t(M_X)$, as discussed earlier, is a rough constraint
for perturbative believability, and if one wished to include higher-loop
perturbative contributions there is a {\it prima facie} case for restricting
the allowed range of $y_t(M_X)$ still further. 
The PB results show the same QIRFP behaviour;
however, we believe they are reliable up to a larger value of $y_t(M_X)$. 
Of course,
there must still be an upper limit on $y_t$ beyond which we can no longer
trust the PB approximation. Some indication of where this upper limit is can
be obtained by a comparison with alternative means of implementing the PB
programme; clearly one can no longer trust the PB results in regions
where different approaches give qualitatively different behaviour. For 
instance, one could write
\eqn\padef{
{{dy_t}\over{dt}}=6y_t^2\left[1-\frak{1}{4\pi}\frak{22}{6}y_t+
\left(\frak{1}{4\pi}\right)^2\frak{1}{6}[102+36\zeta(3)]y_t^2+\ldots\right]}
and then perform the PB procedure on the series $1+ay_t+by_t^2+cy_t^3+\ldots$.
Note that the $L$-loop contribution to the perturbation expansion
corresponds to the $O(y_t^{L-1})$ term in this series. This appears at first
to be a drawback, as the PB process will not now exactly match the 
$L!$ growth of the $L$-loop perturbation theory contribution. 
However we might well expect behaviour of the 
$L$-loop contribution like $L!L^a$, and our uncertainty as regards the value
of $a$ means that we cannot really distinguish between growth like $L!$ or
like $(L-1)!$.
 
%The $[1,1]$ PB approximation for the series
%\eqn\padeg{
%{{dy_t}\over{dt}}=6y_t^2(1+ay_t+by_t^2)}
%is readily computed as
%\eqn\padegg{
%{{dy_t}\over{dt}}=6y_t^2\left[1-{2b^2\over c}+{4b^3\over c^2y_t}
%e^{2b\over{cy_t}}
%{\rm E_1}\left({2b\over{cy_t}}\right)\right],}
%The $[2,1]$ PB approximation for the series
%\eqn\padei{
%{{dy_t}\over{dt}}=6y_t^2(1+ay_t+by_t^2+cy_t^3)}
%is given by
%\eqn\padej{
%{{dy_t}\over{dt}}=6y_t^2\left[{{(3b^2-2ac)y_t}\over{6b}}+{1\over2}
%-{9b^3\over{2c^2}}+{1\over2}\left(1+{9b^3\over c^2}\right) {3b\over{cy_t}}
%e^{3b\over{cy_t}}{\rm E_1}\left({3b\over{cy_t}}\right)\right].}
%Once again, the $[1,2]$ PB is ill-defined.
If one evolves $y_t$ alone (setting $\alpha_{1,2,3}=y_b=y_{\tau}=0$),
using the $[1,1]$ or $[2,1]$ PB obtained in this 
way, one finds a fixed point at large $y_t$, for $y_t\sim 100$ and $y_t\sim 8$
respectively, in contrast with the Landau pole behaviour of the previous PBs.
However, in the presence of the other couplings, the behaviour is 
somewhat modified. In fact, the evolution is 
practically indistinguishable from that for the $[2,2]$ PB displayed in Fig.~3
over the range shown. Moreover, for larger $y_t(M_X)$, the fixed point is
obscured by the development of a Landau pole in $y_b$ and by the increasing
size of $y_t^3\alpha_3$ terms which have not been Borel summed. 
Nevertheless, the last vestige
of a fixed point is discernible in the $[2,1]$ PB evolution at $y_t\sim 6$.  
This may perhaps be taken as a signal that we should not trust any of our PBs 
beyond this point. Nevertheless we have considerably extended the domain of 
attraction of the QIRFP as compared to the perturbative case beyond 
$y_t(M_X)=1$, and perhaps optimistically as far as $y_t(M_X)\sim5$; 
we now see that 
values of $y_t(M_X)$ in the range $0.2<y_t(M_X)<5$ lead to values of 
$y_t(M_Z)$ in the range $0.089<y_t(M_Z)<0.094$.  
Note that the fixed point for the $[2,2]$ PB is roughly 2\% higher than the 
effective upper limit on $y_t(M_Z)$ which applies in the one-loop
perturbative case; for a fixed value of $m_t$, this leads to a reduction in
$\tan \beta$ of around 4.5\%. 

In conclusion, we have demonstrated by means of Pad\'e-Borel summation that
the domain of attraction of the quasi-infra-red
fixed point in the supersymmetric standard model is in fact large. 
There has been some recent 
speculation\ref\IRSU{P.M.~Ferreira, I.~Jack and D.R.T.~Jones,
\plb357(1995) 359\semi
M.~Lanzagorta and G.G.~Ross, \plb364 (1995) 163\semi
J.~Kubo, M.~Mondragon and G.~Zoupanos, hep-ph/9609218}  
on the possible r\^ ole of QIRFP behaviour in the theory above
$M_X$ in assuring universality of the soft-breaking parameters. Because the 
energy range between $M_{\rm Planck}$ and $M_{X}$ is much smaller than
that between $M_X$ and $M_Z$, rapid evolution of the couplings is essential 
if a QIRFP is to be approached in this region. Since the approach to the QIRFP
appears to be quicker for larger initial couplings, it might be of interest
to use PB summation to explore the region of larger coupling with more
confidence. 

\bigskip\centerline{{\bf Acknowledgements}}\nobreak

While part of this work was done, two of us (IJ and TJ) enjoyed the 
hospitality of the Aspen Center for Physics. 
We thank Professor Avdeev for correspondence, and David Barclay for 
conversations. PF was supported by a
scholarship from JNICT.  

\listrefs
\bye